\documentclass[apjl]{emulateapj}
\usepackage{comment}
\usepackage{ifthen}
\usepackage{lineno}
\usepackage{lipsum}

\newcommand{\forloop}[5][1]%
{%
\setcounter{#2}{#3}%
\ifthenelse{#4}%
	{%
	#5%
	\addtocounter{#2}{#1}%
	\forloop[#1]{#2}{\value{#2}}{#4}{#5}%
	}%
	{%
	}%
}%

\newcommand{\ctbd}[1]{}

\newcommand{\lc}{light curve}
\newcommand{\lcs}{light curves}
\newcommand{\Lc}{Light curve}

\newcommand{\band}[1]{\ensuremath{#1}~band}

\newcommand{\kms}{\ensuremath{\rm km\,s^{-1}}}
\newcommand{\ms}{\ensuremath{\rm m\,s^{-1}}}

\newcommand{\gcmc}{\ensuremath{\rm g\,cm^{-3}}}
\newcommand{\ergscmsq}{\ensuremath{\rm erg\,s^{-1}\,cm^{-2}}}

\newcommand{\logg}{\ensuremath{\log{g}}}
\newcommand{\vsini}{\ensuremath{v \sin{i}}}
\newcommand{\feh}{\ensuremath{\rm [Fe/H]}}

\newcommand{\rsun}{\ensuremath{R_\sun}}
\newcommand{\msun}{\ensuremath{M_\sun}}
\newcommand{\lsun}{\ensuremath{L_\sun}}

\newcommand{\rstar}{\ensuremath{R_\star}}
\newcommand{\mstar}{\ensuremath{M_\star}}
\newcommand{\lstar}{\ensuremath{L_\star}}

\newcommand{\teffstar}{\ensuremath{T_{\rm eff\star}}}
\newcommand{\rhostar}{\ensuremath{\rho_\star}}
\newcommand{\loggstar}{\ensuremath{\log{g_{\star}}}}

\newcommand{\rpl}{\ensuremath{R_{p}}}
\newcommand{\mpl}{\ensuremath{M_{p}}}

\newcommand{\rhopl}{\ensuremath{\rho_{p}}}

\newcommand{\arstar}{\ensuremath{a/\rstar}}
\newcommand{\zrstar}{\ensuremath{\zeta/\rstar}}

\newcommand{\rjup}{\ensuremath{R_{\rm J}}}
\newcommand{\mjup}{\ensuremath{M_{\rm J}}}

\newcommand{\reffig}[1]{Fig.~\ref{fig:#1}}

\newcommand{\reffigl}[1]{Figure~\ref{fig:#1}}
\newcommand{\refsecl}[1]{\mbox{Section \ref{sec:#1}}}

\newcommand{\reftabl}[1]{Table~\ref{tab:#1}}

\newcommand{\flwof}{\mbox{FLWO 1.2\,m}}

\newcommand{\hatcurhtr}{HTR265-005}                       
\newcommand{\hatcurCCra}{\ensuremath{06^{\mathrm h}39^{\mathrm m}35.53{\mathrm s}}}                     
\newcommand{\hatcurCCdec}{\ensuremath{25{\arcdeg}28{\arcmin}57.1{\arcsec}}}                     
\newcommand{\hatcurCCtwomass}{2MASS~06393552+2528571}     
\newcommand{\hatcurCCgsc}{GSC~1884-00168}                 
\newcommand{\hatcurCCtassmv}{\ensuremath{13.505\pm0.060}} 
\newcommand{\hatcurCCtassmB}{\ensuremath{14.832\pm0.070}} 
\newcommand{\hatcurCCtassmg}{\ensuremath{14.181\pm0.020}} 
\newcommand{\hatcurCCtassmr}{\ensuremath{12.976\pm0.020}} 
\newcommand{\hatcurCCtassmi}{\ensuremath{12.483\pm0.020}} 
\newcommand{\hatcurCCtwomassJmag}{\ensuremath{11.145\pm0.025}} 
\newcommand{\hatcurCCtwomassHmag}{\ensuremath{10.487\pm0.030}} 
\newcommand{\hatcurCCtwomassKmag}{\ensuremath{10.326\pm0.019}} 
\newcommand{\hatcurLCrprstar}{\ensuremath{0.1572\pm0.0020}} 
\newcommand{\hatcurLCbsq}{\ensuremath{0.550_{-0.015}^{+0.015}}} 
\newcommand{\hatcurLCimp}{\ensuremath{0.741_{-0.011}^{+0.010}}} 
\newcommand{\hatcurLCzeta}{\ensuremath{35.36\pm0.29}}     
\newcommand{\hatcurLCdur}{\ensuremath{0.07488\pm0.00072}} 
\newcommand{\hatcurLCingdur}{\ensuremath{0.02055\pm0.00093}} 
\newcommand{\hatcurLCP}{\ensuremath{3.799847\pm0.000014}} 
\newcommand{\hatcurLCPshort}{\ensuremath{3.7998}}         
\newcommand{\hatcurLCT}{\ensuremath{2456299.30370\pm0.00024}} 
\newcommand{\hatcurSMEiteff}{\ensuremath{4390\pm50}}      
\newcommand{\hatcurSMEizfeh}{\ensuremath{-0.127\pm0.080}} 
\newcommand{\hatcurSMEizfehshort}{\ensuremath{-0.13}}     
\newcommand{\hatcurSMEilogg}{\ensuremath{4.66\pm0.10}}    
\newcommand{\hatcurSMEivsin}{\ensuremath{2.35\pm0.50}}    
\newcommand{\hatcurSMEivmac}{\ensuremath{0.0}}            
\newcommand{\hatcurSMEivmic}{\ensuremath{0.0}}            
\newcommand{\hatcurSMEiiteff}{\ensuremath{4390\pm50}}     
\newcommand{\hatcurSMEiizfeh}{\ensuremath{-0.127\pm0.080}} 
\newcommand{\hatcurSMEiizfehshort}{\ensuremath{-0.127}}   
\newcommand{\hatcurSMEiilogg}{\ensuremath{4.66\pm0.10}}   
\newcommand{\hatcurSMEiivsin}{\ensuremath{2.35\pm0.50}}   
\newcommand{\hatcurTRESgamma}{\ensuremath{30.84\pm0.18}}  
\newcommand{\hatcurLBii}{\ensuremath{0.4324}}             
\newcommand{\hatcurLBiii}{\ensuremath{0.2457}}            
\newcommand{\hatcurLBir}{\ensuremath{0.5964}}             
\newcommand{\hatcurLBiir}{\ensuremath{0.1699}}            
\newcommand{\hatcurISOm}{\ensuremath{0.645\pm0.020}}      
\newcommand{\hatcurISOmshort}{\ensuremath{0.64}}          
\newcommand{\hatcurISOmlong}{\ensuremath{0.645\pm0.020}}  
\newcommand{\hatcurISOr}{\ensuremath{0.617\pm0.013}}      
\newcommand{\hatcurISOrlong}{\ensuremath{0.617\pm0.013}}  
\newcommand{\hatcurISOlogg}{\ensuremath{4.667\pm0.012}}   
\newcommand{\hatcurISOlum}{\ensuremath{0.1254\pm0.0089}}  
\newcommand{\hatcurISOmv}{\ensuremath{7.78\pm0.13}}       
\newcommand{\hatcurISOage}{\ensuremath{3.9_{-2.1}^{+4.3}}} 
\newcommand{\hatcurISOMK}{\ensuremath{4.720\pm0.054}}     
\newcommand{\hatcurRVK}{\ensuremath{132.6\pm4.9}}         
\newcommand{\hatcurRVjitterA}{\ensuremath{3.2\pm1.7}}     
\newcommand{\hatcurRVjitterB}{\ensuremath{53\pm15}}       
\newcommand{\hatcurPPi}{\ensuremath{87.040\pm0.084}}      
\newcommand{\hatcurPPlogg}{\ensuremath{3.324\pm0.027}}    
\newcommand{\hatcurPPar}{\ensuremath{14.34\pm0.22}}       
\newcommand{\hatcurPParel}{\ensuremath{0.04117\pm0.00043}} 
\newcommand{\hatcurPPrho}{\ensuremath{1.118\pm0.098}}     
\newcommand{\hatcurPPm}{\ensuremath{0.760\pm0.032}}       
\newcommand{\hatcurPPmlong}{\ensuremath{0.760\pm0.032}}   
\newcommand{\hatcurPPr}{\ensuremath{0.944\pm0.028}}       
\newcommand{\hatcurPPrlong}{\ensuremath{0.944\pm0.028}}   
\newcommand{\hatcurPPmrcorr}{\ensuremath{0.30}}           
\newcommand{\hatcurPPteff}{\ensuremath{818\pm12}}         
\newcommand{\hatcurPPtheta}{\ensuremath{0.1025\pm0.0050}} 
\newcommand{\hatcurPPfluxavg}{\ensuremath{1.010\pm0.060}} 
\newcommand{\hatcurPPfluxavgdim}{\ensuremath{8}}          
\newcommand{\hatcurXAv}{\ensuremath{0.12\pm0.11}}         
\newcommand{\hatcurXdistred}{\ensuremath{135.8\pm3.5}}    
\newcommand{\hatcurRVeccentwosiglimeccendartmouth}{\ensuremath{<0.074}} 
\newcommand{\hatcur}{HAT-P-54}
\newcommand{\hatcurb}{HAT-P-54b}

\newcommand{\hatcurlumind}{\rhostar}
\newcommand{\hatcurjhkfilset}{ESO}

\newcommand{\hatcurSMEversion}{i}                                       
\newcommand{\hatcurSMEteff}{\ifthenelse{\equal{\hatcurSMEversion}{i}}{\hatcurSMEiteff}{\hatcurSMEiiteff}}
\newcommand{\hatcurSMEzfeh}{\ifthenelse{\equal{\hatcurSMEversion}{i}}{\hatcurSMEizfeh}{\hatcurSMEiizfeh}}
\newcommand{\hatcurSMEzfehshort}{\ifthenelse{\equal{\hatcurSMEversion}{i}}{\hatcurSMEizfehshort}{\hatcurSMEiizfehshort}}
\newcommand{\hatcurSMElogg}{\ifthenelse{\equal{\hatcurSMEversion}{i}}{\hatcurSMEilogg}{\hatcurSMEiilogg}}
\newcommand{\hatcurSMEvsin}{\ifthenelse{\equal{\hatcurSMEversion}{i}}{\hatcurSMEivsin}{\hatcurSMEiivsin}}
\newcommand{\hatcurSMEvmac}{\ifthenelse{\equal{\hatcurSMEversion}{i}}{\hatcurSMEivmac}{\hatcurSMEiivmac}}
\newcommand{\hatcurSMEvmic}{\ifthenelse{\equal{\hatcurSMEversion}{i}}{\hatcurSMEivmic}{\hatcurSMEiivmic}}

\newcommand{\hatcurSMEteffcirc}{\ifthenelse{\equal{\hatcurSMEversion}{i}}{\hatcurSMEiteffcirc}{\hatcurSMEiiteffcirc}}
\newcommand{\hatcurSMEzfehcirc}{\ifthenelse{\equal{\hatcurSMEversion}{i}}{\hatcurSMEizfehcirc}{\hatcurSMEiizfehcirc}}
\newcommand{\hatcurSMEzfehshortcirc}{\ifthenelse{\equal{\hatcurSMEversion}{i}}{\hatcurSMEizfehshortcirc}{\hatcurSMEiizfehshortcirc}}
\newcommand{\hatcurSMEloggcirc}{\ifthenelse{\equal{\hatcurSMEversion}{i}}{\hatcurSMEiloggcirc}{\hatcurSMEiiloggcirc}}
\newcommand{\hatcurSMEvsincirc}{\ifthenelse{\equal{\hatcurSMEversion}{i}}{\hatcurSMEivsincirc}{\hatcurSMEiivsincirc}}
\newcommand{\hatcurSMEvmaccirc}{\ifthenelse{\equal{\hatcurSMEversion}{i}}{\hatcurSMEivmaccirc}{\hatcurSMEiivmaccirc}}
\newcommand{\hatcurSMEvmiccirc}{\ifthenelse{\equal{\hatcurSMEversion}{i}}{\hatcurSMEivmiccirc}{\hatcurSMEiivmiccirc}}

\newcommand{\hatcurSMEteffeccen}{\ifthenelse{\equal{\hatcurSMEversion}{i}}{\hatcurSMEiteffeccen}{\hatcurSMEiiteffeccen}}
\newcommand{\hatcurSMEzfeheccen}{\ifthenelse{\equal{\hatcurSMEversion}{i}}{\hatcurSMEizfeheccen}{\hatcurSMEiizfeheccen}}
\newcommand{\hatcurSMEzfehshorteccen}{\ifthenelse{\equal{\hatcurSMEversion}{i}}{\hatcurSMEizfehshorteccen}{\hatcurSMEiizfehshorteccen}}
\newcommand{\hatcurSMEloggeccen}{\ifthenelse{\equal{\hatcurSMEversion}{i}}{\hatcurSMEiloggeccen}{\hatcurSMEiiloggeccen}}
\newcommand{\hatcurSMEvsineccen}{\ifthenelse{\equal{\hatcurSMEversion}{i}}{\hatcurSMEivsineccen}{\hatcurSMEiivsineccen}}
\newcommand{\hatcurSMEvmaceccen}{\ifthenelse{\equal{\hatcurSMEversion}{i}}{\hatcurSMEivmaceccen}{\hatcurSMEiivmaceccen}}
\newcommand{\hatcurSMEvmiceccen}{\ifthenelse{\equal{\hatcurSMEversion}{i}}{\hatcurSMEivmiceccen}{\hatcurSMEiivmiceccen}}

\newcounter{planetcounter}

\newboolean{emulateapj}
\setboolean{emulateapj}{true}

\newboolean{rvtablelong}
\setboolean{rvtablelong}{true}

\newboolean{astroph}
\setboolean{astroph}{true}

\shortauthors{Bakos et al.}
\shorttitle{
\hatcur\lowercase{b}
}
\ifthenelse{\boolean{emulateapj}}{
    \newcommand{\titledag}{$\dagger$}
}{
    \newcommand{\titledag}{\dagger}
}

\begin{document}
\title{
\hatcur\lowercase{b}: A Hot Jupiter Transiting a
\hatcurISOmshort\,\msun\ Star in Field 0 of the K2 Mission\altaffilmark{\titledag}
}

\author{
    G.~\'A.~Bakos\altaffilmark{1,9,10},
    J.~D.~Hartman\altaffilmark{1},
    W.~Bhatti\altaffilmark{1}, 
    A.~Bieryla\altaffilmark{2},
    M.~de Val-Borro\altaffilmark{1},
    D.~W.~Latham\altaffilmark{2},
    L.~A.~Buchhave\altaffilmark{2,3},
    Z.~Csubry\altaffilmark{1},
    K.~Penev\altaffilmark{1},
    G.~Kov\'acs\altaffilmark{4,5}, 
    B.~B\'eky\altaffilmark{2},
    E.~Falco\altaffilmark{2},
    T.~Kov\'acs\altaffilmark{4},
    A.~W.~Howard\altaffilmark{6},
    J.~A.~Johnson\altaffilmark{2,9,10},
    H.~Isaacson\altaffilmark{7},
    G.~W.~Marcy\altaffilmark{7},
    G.~Torres\altaffilmark{2},
    R.~W.~Noyes\altaffilmark{2},
    P.~Berlind\altaffilmark{2},
    M.~L.~Calkins\altaffilmark{2},
    G.~A.~Esquerdo\altaffilmark{2},
    J.~L\'az\'ar\altaffilmark{8},
    I.~Papp\altaffilmark{8},
    P.~S\'ari\altaffilmark{8}
}

\altaffiltext{1}{Department of Astrophysical Sciences, Princeton
	University, Princeton, NJ 08544 USA; email:
	gbakos@astro.princeton.edu}

\altaffiltext{2}{Harvard-Smithsonian Center for Astrophysics,
    Cambridge, MA 02138 USA; email: abieryla@cfa.harvard.edu}

\altaffiltext{3}{Centre for Star and Planet Formation, Natural History
	Museum of Denmark, University of Copenhagen, DK-1350 Copenhagen,
	Denmark}

\altaffiltext{4}{Konkoly Observatory, Budapest, Hungary}

\altaffiltext{5}{Department of Physics and Astrophysics, University of
	North Dakota, Grand Forks, ND USA}

\altaffiltext{6}{Institute for Astronomy, University of Hawaii,
	Honolulu, HI 96822}

\altaffiltext{7}{Department of Astronomy, University of California,
	Berkeley, CA}

\altaffiltext{8}{Hungarian Astronomical Association (HAA).}

\altaffiltext{9}{Sloan Fellow}

\altaffiltext{10}{Packard Fellow}

\altaffiltext{$\dagger$}{
Based on observations obtained with the Hungarian-made Automated
Telescope Network (HATNet).  Based in part on observations obtained at
the W.~M.~Keck Observatory, using time granted by NASA (N133Hr).  Based
in part on observations obtained with the Tillinghast Reflector 1.5\,m
telescope and the 1.2\,m telescope, both operated by the Smithsonian
Astrophysical Observatory at the Fred Lawrence Whipple Observatory in
Arizona.
}


\begin{abstract}

\setcounter{footnote}{10}
We report the discovery of \hatcurb{}, a planet transiting a late K
dwarf star in field 0 of the NASA K2 mission.  We combine ground-based
photometric light curves with radial velocity measurements to determine
the physical parameters of the system.  \hatcurb{} has a mass of
$\hatcurPPm$\,\mjup, a radius of $\hatcurPPr$\,\rjup, and an orbital
period of $\hatcurLCPshort$\,d.  The star has $V=\hatcurCCtassmv$, a
mass of $\hatcurISOm$\,\msun, a radius of $\hatcurISOr$\,\rsun, an
effective temperature of $\teffstar=\hatcurSMEteff$, and a subsolar
metallicity of $\feh=\hatcurSMEzfeh$.  \hatcurb{} has a radius that is
smaller than 92\% of the known transiting planets with masses greater
than that of Saturn, while \hatcur{} is one of the lowest-mass stars
known to host a hot Jupiter.  Follow-up high-precision photometric
observations by the K2 mission promise to make this a well-studied
planetary system.
\setcounter{footnote}{0}
\end{abstract}



\section{Introduction}
\label{sec:introduction}

Among the best studied transiting planets are the three planets
discovered by wide-field ground-based surveys in the field of the NASA
{\em Kepler} mission prior to the mission launch.  These planets,
including TrES-2b \citep{odonovan:2006}, HAT-P-7b \citep{pal:2008:hat7}
and HAT-P-11b \citep{bakos:2010:hat11} orbit bright stars, have
relatively deep transits, and have short orbital periods.  These same
factors enabled their discovery from the ground.  The extremely high
S/N ratio transit observations by {\em Kepler} have allowed a number of
subtle effects to be studied for TrES-2b and HAT-P-7b.  These include
the optical phase variation and secondary eclipse due to reflected
light from the planet, the tidal distortion of the star due to the
planet, Doppler beaming, and the detection of p-mode oscillations
enabling asteroseismology of the host stars, among others
\citep{morris:2013,barclay:2012,jackson:2012,kipping:2011,welsh:2010,borucki:2009}. 
For HAT-P-11b observations of starspot crossings by the planet have
revealed the presence of active latitudes on the star, have been used
to show that there is a high inclination between the spin axis of the
star and the orbital axis of the planet
\citep{sanchisojeda:2011,deming:2011}, and have also revealed an
apparent commensurability between the orbital period of the planet and
the rotation period of the star \citep{beky:2014}.  While the transits
of these three systems could have easily been discovered from the {\em
Kepler} light curves themselves, the prior ground-based detections
ensured that the targets would be included among the limited number of
stars for which data are downloaded, they extended the time base-line
over which transits have been measured, and they provided a set of
confirmed planets, including radial velocities (RVs) used to determine
their masses, for which the performance of {\em Kepler} could be
ascertained immediately after launch.

Following the failure of two of the {\em Kepler} reaction wheels, a
repurposed {\em Kepler} mission, dubbed K2, has been proposed
\citep{howell:2014}.  In this mission the {\em Kepler} space telescope
will be used to observe 10 fields along the ecliptic plane over the
course of two years.  Due to various constraints the number of stars
that can be observed in each field is substantially lower than the
number for the original {\em Kepler} mission.  In this case prior
observations of the K2 fields by ground-based telescopes to preselect
targets are extremely valuable.

In this paper we present the discovery of a transiting planet,
\hatcurb{}, in the first field that is observed by the K2 mission
(called field 0).  This planet, discovered by the HATNet survey
\citep{bakos:2004:hatnet}, is the first transiting planet to be
identified in this field.

In \refsecl{obs} we summarize the detection of the photometric transit
signal and the subsequent spectroscopic and photometric observations of
the star to confirm the planet.  In \refsecl{analysis} we analyze the
data to rule out false positive scenarios, and to determine the stellar
and planetary parameters.  We discuss our findings briefly in
\refsecl{discussion}.

\section{Observations}
\label{sec:obs}

\subsection{Photometry}
\label{sec:photometry}

\subsubsection{Photometric detection}
\label{sec:detection}

Photometric observations of the star \hatcur{} (see identifying
information in \reftabl{stellar}) were carried out by the fully
automated HATNet system \citep{bakos:2004:hatnet} between 2011 Oct and
2012 May using the HAT-6 instrument at Fred Lawrence Whipple
Observatory (FLWO) in Arizona, and between 2011 Oct and 2012 Feb using
the HAT-9 instrument at Mauna Kea Observatory (MKO) in Hawaii.  A total
of 6609 images yielding acceptable photometry were obtained with HAT-6,
and 4233 images with HAT-9.  We used an exposure time of 180\,s (median
cadence of 214\,s) and a Sloan $r$-band filter for the observations. 
Data were reduced to trend-filtered light curves following
\cite{bakos:2010:hat11}.  The trend filtering included de-correlating
the light curves against a set of measured parameters (which we refer
to as External Parameter Decorrelation; or EPD), followed by
application of the Trend Filtering Algorithm
\citep[TFA;][]{kovacs:2005:TFA}.  The per-point RMS scatter of the
resulting light curves is 23.6\,mmag for HAT-6 and 20.4\,mmag for
HAT-9, and is dominated by shot noise from the sky background.

\begin{figure}[]
\plotone{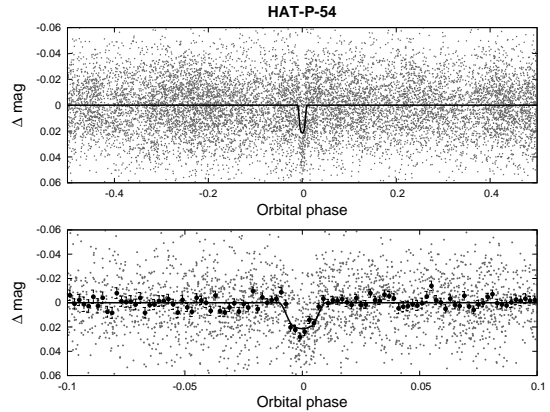}
\caption[]{
    HATNet \lc{} of \hatcur\ phase folded with the transit period.  The
    top panel shows the unbinned light curve, while the bottom shows
    the region zoomed-in on the transit, with dark filled circles for
    the light curve binned in phase with a binsize of 0.002.  The solid
    line shows the model fit to the light curve.
\label{fig:hatnet}}
\end{figure}

Transits were detected in the combined light curve using the
Box-fitting Least Squares algorithm \citep[BLS;][]{kovacs:2002:BLS}. 
The trend-filtered, phase-folded, combined light curve for \hatcur{} is
shown in \reffigl{hatnet}, while the individual photometric
measurements are provided in Table~\ref{tab:phfu}.

We used BLS to search for additional transiting signals in the residual
HATNet light curve (after subtracting the transits of \hatcurb{}) but
found nothing significant above the noise.  However, we found a
significant sinusoidal variation most likely due to stellar activity
(i.e., modulation of the brightness due to spots rotating on the
surface of the star; see \refsecl{rotation} for details).

In the following subsections we discuss the
observations used to confirm \hatcurb{} as a transiting planet.

\subsubsection{Photometric follow-up}
\label{sec:phfu}

\begin{figure}[]
\plotone{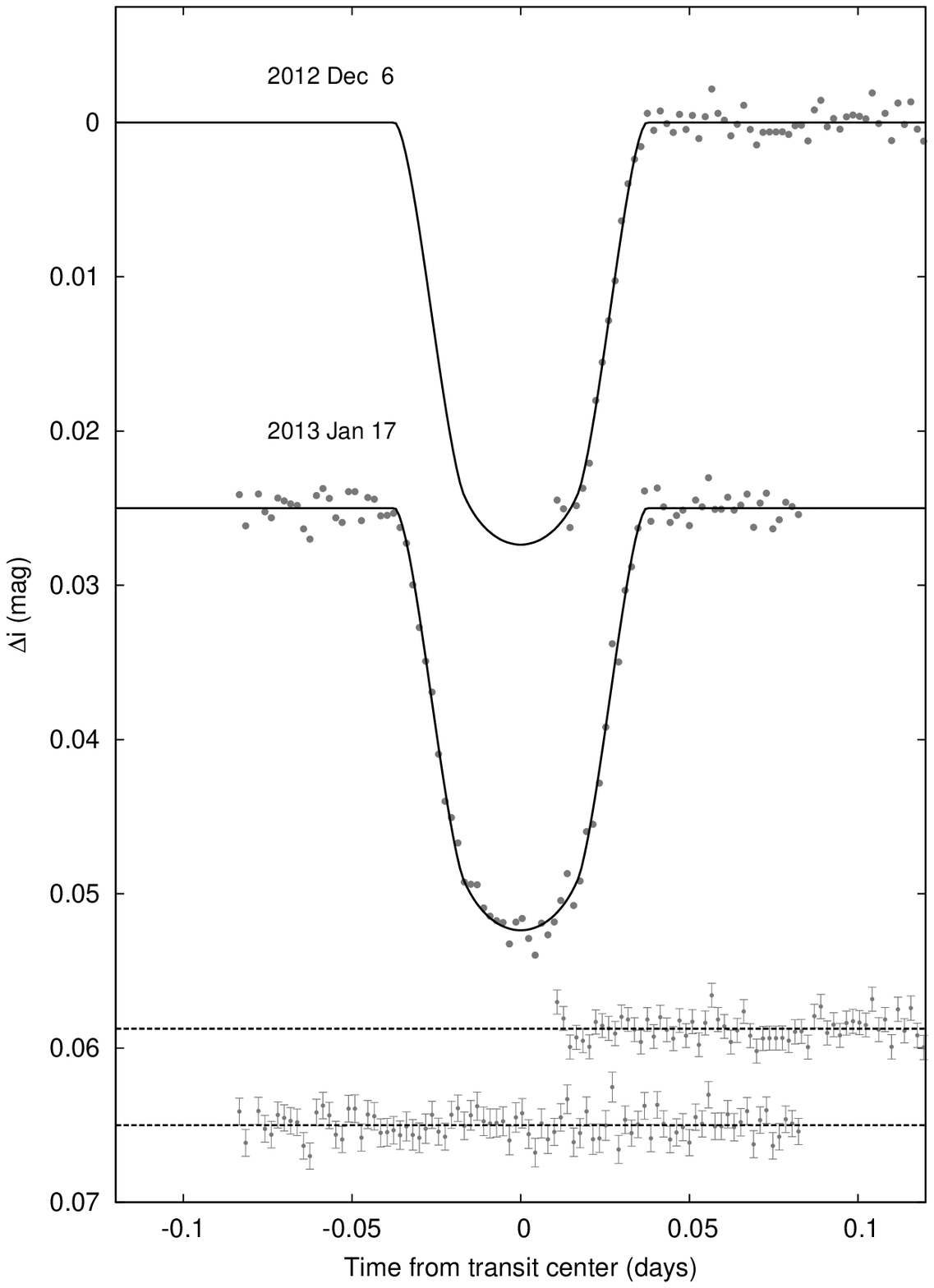}
\caption{
    Unbinned transit \lcs{} for \hatcur, acquired with KeplerCam at the
    \flwof{} telescope.  The light curves have been EPD- and
    TFA-processed, as described in \citet{bakos:2010:hat11}.  The dates
    of each event are indicated.  Our best fit from the global
    modeling described in \refsecl{analysis} is shown by the solid
    line.  Residuals from the fit are displayed below in the same order
    as the original light curves.  The error bars represent the photon
    and background shot noise, plus the readout noise.
}
\label{fig:lc}
\end{figure}

Higher precision photometric time series observations were obtained
using the Keplercam imager on the FLWO~1.2\,m telescope.  We observed
an egress on the night of 2012 Dec 6, and a full transit on the night
of 2013 Jan 17.  A total of 72 observations were made on the first
night, and 87 were made on the second night.  In both cases we used a
Sloan $i$-band filter, and an exposure time of 150\,s (yielding a
median cadence of 165\,s).  The images were reduced to light curves
following \citet{bakos:2010:hat11}.  We corrected the light curves for
systematic variations using the EPD and TFA procedures as part of our
model fitting procedure (Section~\ref{sec:analysis}). 
Figure~\ref{fig:lc} shows the resulting trend-filtered light curves,
together with our best fit transiting planet light curve model.  The
data are provided in Table~\ref{tab:phfu}.  The residuals from the
best-fit model have a per-point RMS scatter of 0.9\,mmag on each night.

\ifthenelse{\boolean{emulateapj}}{
    \begin{deluxetable*}{lrrrrr}
}{
    \begin{deluxetable}{lrrrrr}
}
\tablewidth{0pc}
\tablecaption{
    Differential photometry of
    \hatcur\label{tab:phfu}.
}
\tablehead{
    \colhead{BJD\tablenotemark{a}} & 
    \colhead{Mag\tablenotemark{b}} & 
    \colhead{\ensuremath{\sigma_{\rm Mag}}} &
    \colhead{Mag(orig)\tablenotemark{c}} & 
    \colhead{Filter} &
    \colhead{Instrument} \\
    \colhead{\hbox{~~~~(2,400,000$+$)~~~~}} & 
    \colhead{} & 
    \colhead{} &
    \colhead{} & 
    \colhead{} & 
    \colhead{}
}
\startdata
$ 55848.92474 $ & $   0.03198 $ & $   0.03424 $ & $ \cdots $ & $ r$ &     HATNet\\
$ 55848.92799 $ & $   0.02924 $ & $   0.02342 $ & $ \cdots $ & $ r$ &     HATNet\\
$ 55848.93068 $ & $  -0.02762 $ & $   0.03088 $ & $ \cdots $ & $ r$ &     HATNet\\
$ 55848.93857 $ & $   0.01040 $ & $   0.02630 $ & $ \cdots $ & $ r$ &     HATNet\\
$ 55848.94109 $ & $   0.02453 $ & $   0.02663 $ & $ \cdots $ & $ r$ &     HATNet\\
$ 55848.94365 $ & $  -0.02504 $ & $   0.02325 $ & $ \cdots $ & $ r$ &     HATNet\\
$ 55848.94856 $ & $   0.03380 $ & $   0.02507 $ & $ \cdots $ & $ r$ &     HATNet\\
$ 55848.95633 $ & $   0.03176 $ & $   0.02936 $ & $ \cdots $ & $ r$ &     HATNet\\
$ 55848.95889 $ & $  -0.00495 $ & $   0.02720 $ & $ \cdots $ & $ r$ &     HATNet\\
$ 55848.96144 $ & $  -0.02975 $ & $   0.02380 $ & $ \cdots $ & $ r$ &     HATNet\\

\enddata
\tablenotetext{a}{
    Barycentric Julian Date calculated directly from UTC, {\em
      without} correction for leap seconds.
}
\tablenotetext{b}{
    The out-of-transit level has been subtracted. These magnitudes
    have been subjected to the EPD and TFA procedures, carried out
    simultaneously with the transit fit for the follow-up data. For
    HATNet this filtering was applied {\em before} fitting for the
    transit.
}
\tablenotetext{c}{
    Raw magnitude values after correction using comparison stars, but
    without application of the EPD and TFA procedures. This is only
    reported for the follow-up light curves.
}
\tablecomments{
    This table is available in a machine-readable form in the online
    journal.  A portion is shown here for guidance regarding its form
    and content.
}
\ifthenelse{\boolean{emulateapj}}{
    \end{deluxetable*}
}{
    \end{deluxetable}
}

\subsection{Spectroscopy}
\label{sec:hispec}

We carried out optical spectroscopic observations of \hatcur{} using
the Tillinghast Reflector Echelle Spectrograph
\citep[TRES;][]{furesz:2008} on the Tillinghast Reflector 1.5\,m
telescope at FLWO, and the HIRES spectrograph \citep{vogt:1994} on the
Keck-I 10\,m telescope at MKO.  A total of 14 TRES spectra were
obtained using the medium resolution fiber on nights between 2012 Oct
28 and 2013 Nov 14, while 4 HIRES spectra, including an I$_{2}$-free
template spectrum and 3 exposures with the I$_{2}$ cell, were obtained
on the nights of 2013 Oct 18 and 19, and 2013 Dec 12.

\begin{figure} []
\plotone{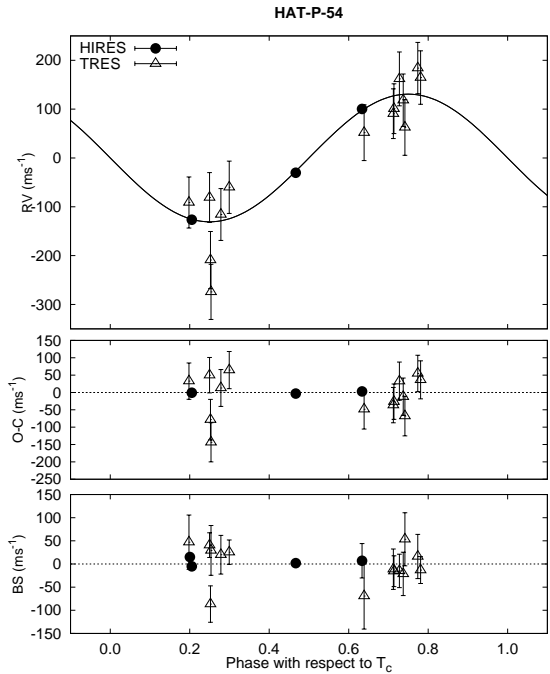}
\caption{
    {\em Top panel:} RV measurements from Keck-I/HIRES (filled
    circles) and FLWO~1.5\,m/TRES (open triangles) for
    \hbox{\hatcur} shown as a function of orbital phase, along with
    our best-fit circular model (solid line; see
    \reftabl{planetparam}).  
    Zero phase corresponds to the time of mid-transit.  The
    center-of-mass velocity has been subtracted.
    {\em Second panel:} Velocity $O\!-\!C$ residuals from the best
    fit. The error bars include a ``jitter'' component
    (\hatcurRVjitterA\,\ms, and \hatcurRVjitterB\,\ms\ for HIRES and
    TRES respectively) added in quadrature to the formal errors (see
    \refsecl{hispec}). The symbols are as in the upper panel.
    {\em Third panel:} Bisector spans (BS), with the mean value
    subtracted. For Keck/HIRES the measurement from the template
    spectrum is included.
    Note the different vertical scales of the panels.
}
\label{fig:rvbis}
\end{figure}

The first two TRES observations were obtained at orbital phases of 0.74
and 0.25 (where phase 0 refers to the center of the transit) so as to
efficiently rule out an eclipsing binary false positive.  Subsequent
observations were also clustered near these phases to maximize our
sensitivity to a planet-induced orbital variation.  We used exposure
times ranging from 2400\,s to 3000\,s yielding a median S/N per
resolution element (SNRe) of 26 near the Mg b region of the spectrum. 
The TRES observations were reduced to RVs and spectral line bisector
spans (BSs) following \cite{buchhave:2010:hat16} and to measurements of
the stellar atmospheric parameters ($\teffstar$, $\logg$, [Fe/H] and
$\vsini$) using the Stellar Parameter Classification (SPC) program
\citep{buchhave:2012}.  The RVs, obtained by conducting an
order-by-order cross correlation against the strongest observed
spectrum, show the predicted sense of variation in phase with the
transit ephemeris, and with a semiamplitude of \hatcurRVK\,\ms\
(Figure~\ref{fig:rvbis}).  The RMS scatter of the TRES residual RVs
from our best-fit circular orbit model is 61\,\ms.  Our model includes
jitter in the amount of \hatcurRVjitterB\,\ms\ which is added in
quadrature to the uncertainties output by the reduction pipeline.  This
jitter term is varied in the fit following \citet{bieryla:2014:hat49}. 
The TRES RV residuals exhibit no evidence for a significant trend. 
Such a trend, if it were present, may indicate additional components
(stellar or planetary) in the system.

The stellar atmospheric parameters derived from the TRES spectra, and
listed in Table~\ref{tab:stellar}, indicate that the star is a cool
($\teffstar=\hatcurSMEteff$\,K), slowly rotating
($\vsini=\hatcurSMEvsin$\,\kms), low-metallicity
([Fe/H]$=\hatcurSMEzfeh$), dwarf ($\logg=\hatcurSMElogg$, in cgs
units).  The errors listed are our estimates of the systematic
uncertainties, based on observations of spectroscopic standard stars. 
Note that these uncertainties are general values adopted for the SPC
program as applied to TRES, and do not include additional errors that
may be present for cool stars.  This issue is discussed further at the
end of this subsection.  For all four parameters, the scatter over the
14 observations is less than the estimated systematic uncertainty. 
Taken together, the RVs, light curves, and stellar parameters strongly
indicate that this is a transiting planet system.

The HIRES observations were reduced to relative RV measurements in the
barycentric frame following the procedure of \cite{butler:1996}, and to
BS measurements following \cite{torres:2007:hat3}.  The latter
measurements were corrected for sky contamination following
\cite{hartman:2011:hat18hat19}.  We note that the contamination was
quite significant for the final HIRES observation, for which the
resulting BS uncertainty is $> 30$\,\ms (shown in \reffigl{rvbis}). 
When only 3 I$_{2}$-cell observations are available, the RV pipeline
underestimates the errors, so we assumed an RV uncertainty of $5$\,\ms,
which is typical for HIRES observations with a similar S/N.  We note,
however, that high stellar activity may induce RV jitter that is larger
than this.  As for TRES, we include a jitter term in the model which is
varied in the fit.  Our modeling yields a jitter for the HIRES
observations of \hatcurRVjitterA\,\ms.  As a consistency check on our
atmospheric parameters, we also applied SPC to the I$_{2}$-free
template HIRES spectrum of \hatcur{}.  The values, listed in
Table~\ref{tab:stellar}, are remarkably similar to the mean parameters
derived from the TRES spectra.

One note of caution regarding the atmospheric parameters is that the
SPC results, which rely on synthetic spectra calculated from Kurucz
model atmospheres \citep{castelli:2003,castelli:2004}, are known to be
unreliable for stars with $\teffstar < 4700$\,K.  To mitigate this
problem a prior on the gravity from the $Y^{2}$ isochrones
\citep{yi:2001} is adopted for cool stars which fixes the gravity to
the range allowed by the stellar models for the initial temperature and
metallicity guesses.  For \hatcur{}, running the analysis without
imposing a prior on the gravity yields $\teffstar = 4330$\,K and $\feh
= -0.12$, which is similar to the values found when the prior is used. 
While this, together with the similar results for the TRES and HIRES
spectra, indicates that SPC is consistently finding the same parameters
for the system, regardless of the spectroscopic instrument used or the
manner in which the surface gravity is treated, due to the known
systematic errors in the models we cannot claim with confidence that
the parameters are accurate to within the stated uncertainties.  A
reanalysis using models that are more suitable for cool stars may yield
parameter values that differ significantly from those presented here.

\setcounter{planetcounter}{1}
%

\ifthenelse{\boolean{emulateapj}}{
    \begin{deluxetable*}{lrrrrrrrrrr}
}{
    \begin{deluxetable}{lrrrrrrrrrr}
}
\tablewidth{0pc}
\tablecaption{
    Relative radial velocities, bisector span measurements and stellar
    atmospheric parameters of \hatcur.
    \label{tab:rvs}
}
\tablehead{
    \colhead{BJD\tablenotemark{a}} &
    \colhead{RV\tablenotemark{b}} &
    \colhead{\ensuremath{\sigma_{\rm RV}}\tablenotemark{c}} &
    \colhead{BS} &
    \colhead{\ensuremath{\sigma_{\rm BS}}} &
    \colhead{SNRe} &
    \colhead{\teffstar\tablenotemark{d}} &
    \colhead{\feh\tablenotemark{d}} &
    \colhead{\vsini\tablenotemark{d}} &
    \colhead{Phase} &
    \colhead{Instrument}\\
    \colhead{\hbox{(2,454,000$+$)}} &
    \colhead{(\ms)} &
    \colhead{(\ms)} &
    \colhead{(\ms)} &
    \colhead{(\ms)} &
    \colhead{} &
    \colhead{(K)} &
    \colhead{} &
    \colhead{(\kms)} &
    \colhead{} &
    \colhead{}
}
\startdata
$ 2229.92435 $ & $    62.59 $ & $    31.41 $ & $    53.40 $ & $    57.30 $ &  $25.5$ & $4427$ & $-0.15$ & $2.4$ & $   0.742 $ & TRES \\
$ 2231.86953 $ & $  -274.31 $ & $    30.20 $ & $    29.30 $ & $    53.80 $ &  $24.8$ & $4458$ & $-0.12$ & $3.1$ & $   0.253 $ & TRES \\
$ 2324.81559 $ & $   100.92 $ & $    17.91 $ & $   -15.60 $ & $    33.30 $ &  $24.2$ & $4370$ & $-0.13$ & $1.8$ & $   0.714 $ & TRES \\
$ 2347.70329 $ & $   118.18 $ & $    24.85 $ & $   -21.40 $ & $    46.80 $ &  $27.8$ & $4361$ & $-0.07$ & $2.7$ & $   0.737 $ & TRES \\
$ 2349.65094 $ & $   -80.99 $ & $    18.20 $ & $    40.70 $ & $    26.70 $ &  $30.5$ & $4407$ & $-0.18$ & $2.1$ & $   0.250 $ & TRES \\
$ 2351.64202 $ & $   184.10 $ & $    21.93 $ & $    16.20 $ & $    47.80 $ &  $26.6$ & $4402$ & $-0.15$ & $2.1$ & $   0.774 $ & TRES \\
$ 2353.63889 $ & $   -60.05 $ & $    24.40 $ & $    25.50 $ & $    26.30 $ &  $26.1$ & $4353$ & $-0.21$ & $2.1$ & $   0.299 $ & TRES \\
$ 2377.72779 $ & $    51.84 $ & $    31.48 $ & $   -69.20 $ & $    71.40 $ &  $27.1$ & $4350$ & $0.01$ & $3.1$  & $   0.639 $ & TRES \\
$ 2385.66489 $ & $   161.96 $ & $    27.61 $ & $   -14.90 $ & $    36.10 $ &  $22.1$ & $4393$ & $-0.19$ & $2.1$ & $   0.728 $ & TRES \\
$ 2389.66737 $ & $   164.67 $ & $    26.78 $ & $   -13.20 $ & $    29.00 $ &  $28.0$ & $4348$ & $-0.12$ & $2.2$ & $   0.781 $ & TRES \\
$ 2402.65190 $ & $   -91.17 $ & $    21.40 $ & $    47.10 $ & $    58.60 $ &  $24.1$ & $4397$ & $-0.07$ & $2.6$ & $   0.198 $ & TRES \\
$ 2573.95028 $ & $  -115.85 $ & $    23.33 $ & $    20.00 $ & $    41.80 $ &  $27.1$ & $4405$ & $-0.18$ & $2.2$ & $   0.278 $ & TRES \\
$ 2585.05331 $\tablenotemark{e} & \nodata      & \nodata      & $    15.05 $ & $     4.34 $ & \nodata & $4388$ & $-0.15$ & \nodata & $   0.200 $ & HIRES \\
$ 2585.07169 $ & $  -126.38 $ & $     5.00 $ & $    -5.19 $ & $     2.28 $ & \nodata & \nodata & \nodata & \nodata & $   0.205 $ & HIRES \\
$ 2586.06592 $ & $   -30.30 $ & $     5.00 $ & $     1.82 $ & $     3.09 $ & \nodata & \nodata & \nodata & \nodata & $   0.467 $ & HIRES \\
$ 2605.99848 $ & $    90.67 $ & $    17.91 $ & $   -11.20 $ & $    43.60 $ &  $29.3$ & $4371$ & $-0.15$ & $2.1$ & $   0.712 $ & TRES \\
$ 2611.84876 $ & $  -209.07 $ & $    33.80 $ & $   -86.60 $ & $    39.30 $ &  $24.9$ & $4416$ & $-0.07$ & $2.3$ & $   0.252 $ & TRES \\
$ 2639.89731 $ & $   100.34 $ & $     5.00 $ & $     6.94 $ & $    37.12 $ & \nodata & \nodata & \nodata & \nodata & $   0.634 $ & HIRES \\

\enddata
\tablenotetext{a}{
    Barycentric Julian Date calculated directly from UTC, {\em
      without} correction for leap seconds.
}
\tablenotetext{b}{
    The zero-point of these velocities is arbitrary. An overall offset
    $\gamma_{\rm rel}$ fitted to these velocities in \refsecl{analysis}
    has {\em not} been subtracted. 
}
\tablenotetext{c}{
    Internal errors excluding the component of astrophysical jitter
    considered in \refsecl{analysis}.
}
\tablenotetext{d}{
    Spectroscopic parameters measured from the individual TRES
    spectra, and from the HIRES I$_{2}$-free template spectrum. The
    uncertainties are $\sim 50$\,K, $0.08$\,dex and $0.5$\,\kms\ on
    $\teffstar$, $\feh$ and $\vsini$, respectively.
}
\tablenotetext{e}{
    This is an I$_{2}$-free template spectrum, for which no velocity is
	measured, but BS values are determined.
}
\ifthenelse{\boolean{rvtablelong}}{
}{
} 
\ifthenelse{\boolean{emulateapj}}{
    \end{deluxetable*}
}{
    \end{deluxetable}
}

\section{Analysis}
\label{sec:analysis}

\subsection{Rejecting Blends}
\label{sec:blend}

To rule out blend scenarios that could potentially explain the
observations of \hatcur{} we conducted an analysis similar to that done
in \cite{hartman:2011:hat32hat33,hartman:2012:hat39hat41}.  We find
that although model blended eclipsing binary systems can fit the
available light curves, absolute photometry, and stellar atmospheric
parameters, these models predict BS variations and RV variations that
are inconsistent with the data (BS variations that are 200\,\ms\ or
greater, and RV variations of 1\,\kms\ or greater).  We conclude that
\hatcur{} is a transiting planet system, and not a blended stellar
eclipsing binary system.  We cannot, however, rule out the possibility
that \hatcur{} is a transiting planet orbiting one component of a
binary star.  High resolution imaging or continued RV monitoring is
needed to check for stellar multiplicity \citep[e.g.][]{adams:2013}. 
We caution that dilution from such a companion could alter the inferred
planetary parameters.  For the rest of this paper we assume that
\hatcur{} is an isolated star.

\subsection{Determining Planetary and Stellar Parameters}
\label{sec:planetparam}

We analyzed the system following \cite{bakos:2010:hat11} and
\cite{hartman:2012:hat39hat41}.  We adopted the stellar atmospheric
parameters obtained by applying the SPC program \citep{buchhave:2012}
to the TRES spectra of \hatcur{} (see Section~\ref{sec:hispec}).  These
values were used to determine fixed limb darkening coefficients taken
from the \cite{claret:2004} tabulations.  We simultaneously modeled
the RVs and light curves using an empirical noise model (EPD+TFA) to
account for systematic variations in the Keplercam data.  The modeling
was done through a Differential Evolution Markov-Chain Monte Carlo
procedure \citep{terbraak:2006,eastman:2013} to explore the fitness
landscape and determine the correlation between parameters.  To speed
up the process, we used standard linear algebra methods to optimize the
parameters associated with our light curve noise model at each step in
the Markov-Chain, rather than exploring their distribution as we do for
the parameters used in the physical model.  For the HIRES and TRES RV
data we included a jitter term, added in quadrature to the formal
errors, and varied in the fit as in \cite{hartman:2012:hat39hat41}, we
also treated the zero-points of the two instruments as free and
independent parameters in the fit.  At each point in the resulting
Markov-Chain we determined the stellar density from the fitted
parameters, and used it, together with values for \teffstar\ and \feh\
drawn from Normal distributions with mean and standard deviations set
to the SPC values, to determine the mass, radius, age and luminosity of
\hatcur{} from theoretical stellar evolution models.

\begin{figure*}[]
\plottwo{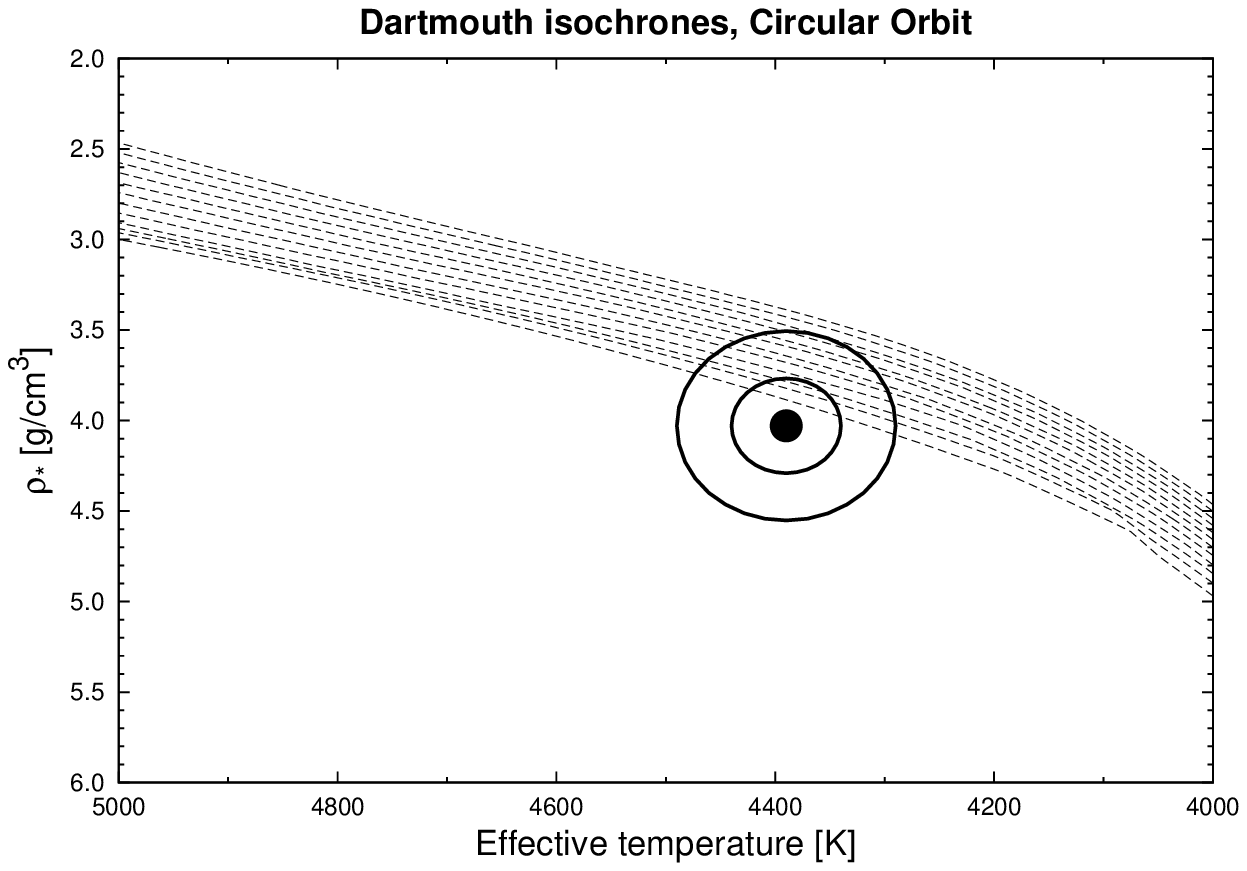}{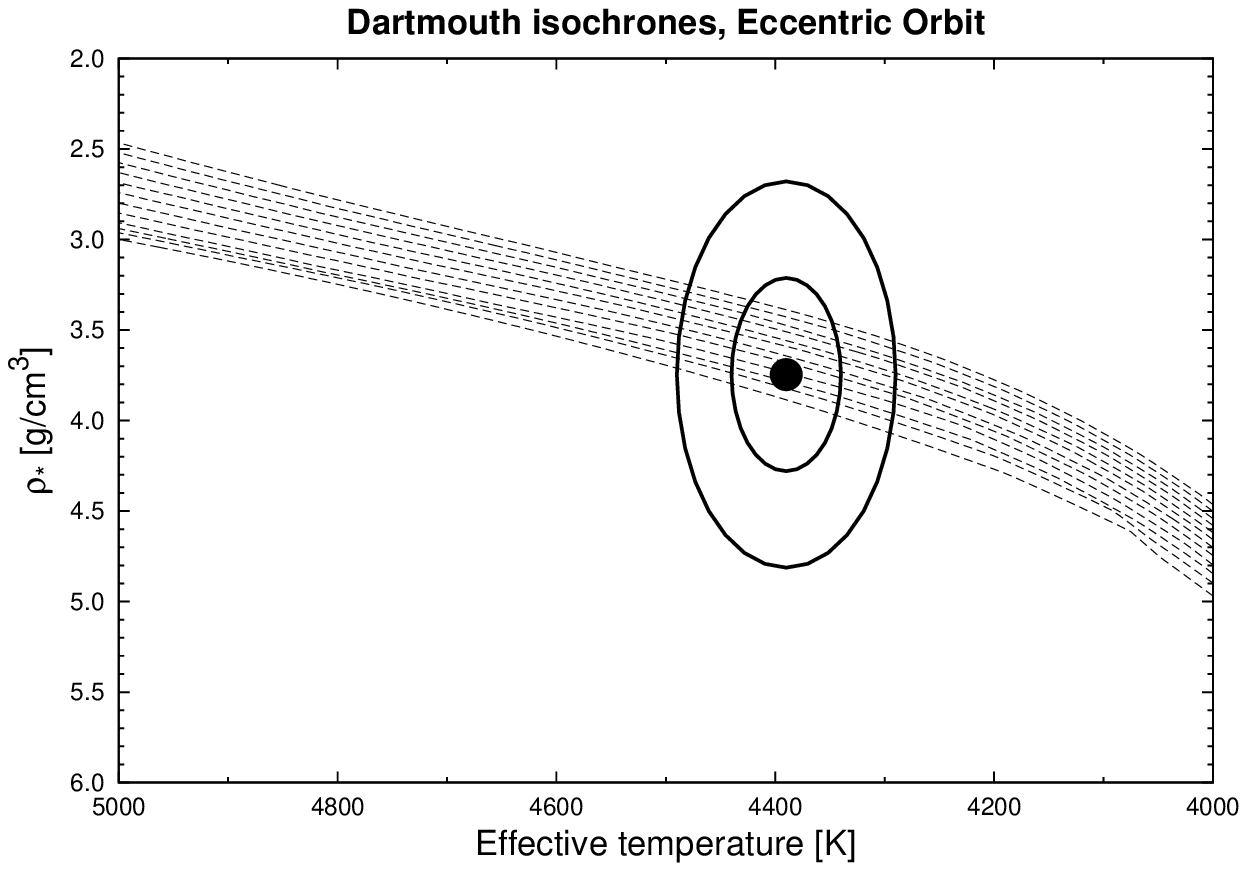}
\caption[]{
    Comparison between the measured values of \teffstar\ and \rhostar\
    (from SPC applied to the TRES spectra, and from our modeling of
    the light curves and RV data, respectively), and model isochrones
    taken from the Dartmouth series \citep{dotter:2008}.  The best-fit
    values, and approximate 1$\sigma$ and 2$\sigma$ confidence
    ellipsoids are shown.  In the left panel we show the measured
    values assuming a circular orbit in the fit, while in the right
    panel we show the values when the eccentricity is allowed to vary
    in the fit.  The isochrones are shown for ages of 1.0 to 13.0\,Gyr
    in 1\,Gyr increments (this is the range over which the models are
    calculated).  In both cases we fix the metallicity to $\feh =
    \hatcurSMEzfehshort$ for display purposes.
\label{fig:iso}}
\end{figure*}

We carried out the analysis both fixing the eccentricity to zero and
allowing it to vary.  We used the Dartmouth \citep{dotter:2008} stellar
evolution models to determine the stellar parameters.  In this respect
we differ from prior HATNet discovery papers which used the Y$^{2}$
\citep{yi:2001} isochrones, which are not optimal for low mass stars
such as \hatcur{}.  Figure~\ref{fig:iso} compares the measured
effective temperature and stellar density for the fixed-circular, and
free-eccentricity models to Dartmouth models.  Because the star is a
late K dwarf, its main sequence lifetime is greater than the age of the
universe.  Assuming the star must be less than 13.8\,Gyr in age
significantly restricts the range of density-temperature-metallicity
combinations permitted by the stellar models.  If the stellar evolution
models are accurate, this effect puts a strong constraint on the
eccentricity of the orbit \citep[c.f.][]{hartman:2011:hat26}.

When the eccentricity is fixed to zero, the combination of median
density, best-fit metallicity, and best-fit temperature falls in a
region of parameter space that is excluded by the Dartmouth models.  In
the \teffstar-\rhostar\ plane, with the metallicity fixed to the
adopted value for the system, the observation falls at a density that
is too high compared to the Dartmouth models.  The observations are,
however, within 1$\sigma$ of the permitted range.  We also note that
the minimum age included in the model is 1\,Gyr, at which time the star
is expected to be already slightly evolved (and thus lower density)
than at the zero-age main sequence.  When the eccentricity is allowed
to vary the inferred stellar density is lower than in the
fixed-circular case bringing the observations into good agreement with
the Dartmouth models.  We used the \cite{weinberg:2013} algorithm to
estimate the Bayesian evidence for both the fixed circular and
free-eccentricity models.  In doing this we restrict the Markov Chains
to consider only those links that are compatible with the stellar
evolution models, and we include $\teffstar$ and \feh\ in calculating
the fitness of each link.  We find that, due to having fewer free
parameters, the fixed-circular model is preferred over the
free-eccentricity model (the evidence ratio is $\sim 60$).  This does
not mean that the orbit is circular, rather the number of
high-precision RV observations is insufficient to place a strong
constraint on the eccentricity.  The 95\% confidence upper-limit based
on the free-eccentricity model is $e
\hatcurRVeccentwosiglimeccendartmouth{}$, with the primary constraint
being the requirement that the observations match to a stellar
evolution model.  If we do not require a match to the stellar evolution
models, then the 95\% confidence upper-limit on the eccentricity is $e
< 0.125$, with the constraint in this case coming only from the RVs.

For our final adopted parameters we use the fixed circular orbit model
due to its higher Bayesian evidence.  Additional high precision RV
observations are necessary to robustly determine the eccentricity of
this system.  The chain of planetary and stellar parameters from our
MCMC analysis is used to estimate the median parameter values together
with their 68.3\% (1$\sigma$) confidence intervals.  These are listed
in Tables~\ref{tab:stellar} and~\ref{tab:planetparam}.  Note that
because the values listed in the table are determined by calculating
the parameter values at each link in the Markov-Chain and then taking
their medians over the Chain, rather than adopting a self-consistent
set of values associated with a particular model, slight numerical
inconsistencies may be apparent when comparing different parameters in
the table.  Although this effect is well-known when presenting
parameters based on a Bayesian analysis, we mention it here to ensure a
proper interpretation of the values listed in the tables.  Assuming a
circular orbit, we conclude that the planet has a mass of
$\hatcurPPmlong$\,\mjup, and a radius of $\hatcurPPrlong$\,\rjup, while
the star has a mass of $\hatcurISOmlong$\,\msun, and a radius of
$\hatcurISOrlong$\,\rsun.

\ifthenelse{\boolean{emulateapj}}{
  \begin{deluxetable*}{lcr}
}{
  \begin{deluxetable}{lcr}
}
\tablewidth{0pc}
\tabletypesize{\scriptsize}
\tablecaption{
    Stellar Parameters for \hatcur{} 
    \label{tab:stellar}
}
\tablehead{
    \multicolumn{1}{c}{~~~~~~~~Parameter~~~~~~~~} &
    \multicolumn{1}{c}{Value\tablenotemark{a}}    &
    \multicolumn{1}{c}{Source}    
}
\startdata
\noalign{\vskip -3pt}
\sidehead{Identifying Information}
~~~~R.A. (h:m:s)                      &  \hatcurCCra{} & 2MASS\\
~~~~Dec. (d:m:s)                      &  \hatcurCCdec{} & 2MASS\\
~~~~GSC ID                            &  \hatcurCCgsc{} & GSC\\
~~~~2MASS ID                          &  \hatcurCCtwomass{} & 2MASS\\
~~~~EPID ID                           &  202126849 & \\
\sidehead{Spectroscopic properties}
~~~~$\teffstar$ (K)\dotfill         &  \hatcurSMEteff{} & TRES+SPC\tablenotemark{b}\\
~~~~$\feh$\dotfill                  &  \hatcurSMEzfeh{} & TRES+SPC                 \\
~~~~$\vsini$ (\kms)\dotfill         &  \hatcurSMEvsin{} & TRES+SPC                 \\
~~~~$\gamma_{\rm RV}$ (\kms)\dotfill&  \hatcurTRESgamma{} & TRES                  \\
\sidehead{Photometric properties}
~~~~$B$ (mag)\dotfill               &  \hatcurCCtassmB{} & APASS                \\
~~~~$V$ (mag)\dotfill               &  \hatcurCCtassmv{} & APASS               \\
~~~~$g$ (mag)\dotfill               &  \hatcurCCtassmg{} & APASS                \\
~~~~$r$ (mag)\dotfill               &  \hatcurCCtassmr{} & APASS                \\
~~~~$i$ (mag)\dotfill               &  \hatcurCCtassmi{} & APASS                \\
~~~~$J$ (mag)\dotfill               &  \hatcurCCtwomassJmag{} & 2MASS           \\
~~~~$H$ (mag)\dotfill               &  \hatcurCCtwomassHmag{} & 2MASS           \\
~~~~$K_s$ (mag)\dotfill             &  \hatcurCCtwomassKmag{} & 2MASS           \\
~~~~$P_{\rm rot}$ (d) \tablenotemark{c}\dotfill         &  $15.6$ & HATNet \\
\sidehead{Derived properties}
~~~~$\mstar$ ($\msun$)\dotfill      &  \hatcurISOmlong{} & Isochrones+\hatcurlumind{}+SPC\tablenotemark{d}\\
~~~~$\rstar$ ($\rsun$)\dotfill      &  \hatcurISOrlong{} & Isochrones+\hatcurlumind{}+SPC         \\
~~~~$\loggstar$ (cgs)\dotfill       &  \hatcurISOlogg{} & Isochrones+\hatcurlumind{}+SPC         \\
~~~~$\lstar$ ($\lsun$)\dotfill      &  \hatcurISOlum{} & Isochrones+\hatcurlumind{}+SPC         \\
~~~~$M_V$ (mag)\dotfill             &  \hatcurISOmv{} & Isochrones+\hatcurlumind{}+SPC         \\
~~~~$M_K$ (mag,\hatcurjhkfilset{})&  \hatcurISOMK{} & Isochrones+\hatcurlumind{}+SPC         \\
~~~~Age (Gyr)\dotfill               &  \hatcurISOage{} & Isochrones+\hatcurlumind{}+SPC         \\
~~~~$A_{V}$ (mag) \tablenotemark{e}\dotfill           &  \hatcurXAv{} & Isochrones+\hatcurlumind{}+SPC\\
~~~~Distance (pc)\dotfill           &  \hatcurXdistred{} & Isochrones+\hatcurlumind{}+SPC\\
\enddata
\tablenotetext{a}{
    The adopted parameters are taken from a model assuming a circular
    orbit and using the Dartmouth isochrones. See
    \refsecl{analysis}. For each parameter with ``Isochrones'' listed
    in the source we give the median value and 68.3\% (1$\sigma$)
    confidence intervals from the MCMC posterior distribution.
}
\tablenotetext{b}{
    SPC = ``Stellar Parameter Classification'' method based on
    cross-correlating high-resolution spectra against synthetic
    templates \citep{buchhave:2012}.}
\tablenotetext{c}{
    Photometric rotation period measured from the HATNet EPD light curve.
}
\tablenotetext{d}{
    Isochrones+\hatcurlumind{}+SPC = Based on the Dartmouth isochrones
    \citep{dotter:2008}, the stellar density used as a luminosity
    indicator, and the SPC results.
} 
\tablenotetext{e}{ Total \band{V} extinction to the star determined
  by comparing the catalog broad-band photometry listed in the table
  to the expected magnitudes from the
  Isochrones+\hatcurlumind{}+SPC model for the star. We use the
  \citet{cardelli:1989} extinction law.}
\ifthenelse{\boolean{emulateapj}}{
  \end{deluxetable*}
}{
  \end{deluxetable}
}
\ifthenelse{\boolean{emulateapj}}{
  \begin{deluxetable*}{lc}
}{
  \begin{deluxetable}{lc}
}
\tabletypesize{\scriptsize}
\tablecaption{Parameters for the transiting planet \hatcurb{}.\label{tab:planetparam}}
\tablehead{
    \multicolumn{1}{c}{~~~~~~~~Parameter~~~~~~~~} &
    \multicolumn{1}{c}{Value \tablenotemark{a}}                     
}
\startdata
\noalign{\vskip -3pt}
\sidehead{\Lc{} parameters}
~~~$P$ (days)             \dotfill    & $\hatcurLCP{}$              \\
~~~$T_c$ (${\rm BJD}$)    
      \tablenotemark{b}   \dotfill    & $\hatcurLCT{}$              \\
~~~$T_{14}$ (days)
      \tablenotemark{b}   \dotfill    & $\hatcurLCdur{}$            \\
~~~$T_{12} = T_{34}$ (days)
      \tablenotemark{b}   \dotfill    & $\hatcurLCingdur{}$         \\
~~~$\arstar$              \dotfill    & $\hatcurPPar{}$             \\
~~~$\zrstar$ \tablenotemark{c}              \dotfill    & $\hatcurLCzeta{}$\phn       \\
~~~$\rpl/\rstar$          \dotfill    & $\hatcurLCrprstar{}$        \\
~~~$b^2$                  \dotfill    & $\hatcurLCbsq{}$            \\
~~~$b \equiv a \cos i/\rstar$
                          \dotfill    & $\hatcurLCimp{}$           \\
~~~$i$ (deg)              \dotfill    & $\hatcurPPi{}$\phn         \\

\sidehead{Limb-darkening coefficients \tablenotemark{d}}
~~~$c_1,i$ (linear term)  \dotfill    & $\hatcurLBii{}$            \\
~~~$c_2,i$ (quadratic term) \dotfill  & $\hatcurLBiii{}$           \\
~~~$c_1,r$               \dotfill    & $\hatcurLBir{}$             \\
~~~$c_2,r$               \dotfill    & $\hatcurLBiir{}$            \\

\sidehead{RV parameters}
~~~$K$ (\ms)              \dotfill    & $\hatcurRVK{}$\phn\phn      \\
~~~$e$ \tablenotemark{e}  \dotfill    & $\hatcurRVeccentwosiglimeccendartmouth{}$ \\
~~~RV jitter Keck-I/HIRES (\ms) \tablenotemark{f}        \dotfill    & \hatcurRVjitterA{}           \\
~~~RV jitter FLWO~1.5\,m/TRES (\ms)        \dotfill    & \hatcurRVjitterB{}           \\

\sidehead{Planetary parameters}
~~~$\mpl$ ($\mjup$)       \dotfill    & $\hatcurPPmlong{}$          \\
~~~$\rpl$ ($\rjup$)       \dotfill    & $\hatcurPPrlong{}$          \\
~~~$C(\mpl,\rpl)$
    \tablenotemark{g}     \dotfill    & $\hatcurPPmrcorr{}$         \\
~~~$\rhopl$ (\gcmc)       \dotfill    & $\hatcurPPrho{}$            \\
~~~$\log g_p$ (cgs)       \dotfill    & $\hatcurPPlogg{}$           \\
~~~$a$ (AU)               \dotfill    & $\hatcurPParel{}$          \\
~~~$T_{\rm eq}$ (K) \tablenotemark{h}        \dotfill   & $\hatcurPPteff{}$           \\
~~~$\Theta$ \tablenotemark{i} \dotfill & $\hatcurPPtheta{}$         \\
~~~$\langle F \rangle$ ($10^{\hatcurPPfluxavgdim}$\ergscmsq) \tablenotemark{i}
                          \dotfill    & $\hatcurPPfluxavg{}$       \\ [-1.5ex]
\enddata
\tablenotetext{a}{
    The adopted parameters are taken from a model assuming a circular
    orbit and using the Dartmouth isochrones. See
    \refsecl{analysis}. For each parameter we give the median value
    and 68.3\% (1$\sigma$) confidence intervals from the MCMC posterior
    distribution.
}
\tablenotetext{b}{
    Reported times are in Barycentric Julian Date calculated directly
    from UTC, {\em without} correction for leap seconds.
    \ensuremath{T_c}: Reference epoch of mid transit that
    minimizes the correlation with the orbital period.
    \ensuremath{T_{14}}: total transit duration, time
    between first to last contact;
    \ensuremath{T_{12}=T_{34}}: ingress/egress time, time between first
    and second, or third and fourth contact.
}
\tablenotetext{c}{
    Reciprocal of the half duration of the transit used as a jump
    parameter in our MCMC analysis in place of $\arstar$. It is
    related to $\arstar$ by the expression $\zrstar = \arstar
    (2\pi(1+e\sin \omega))/(P \sqrt{1 - b^{2}}\sqrt{1-e^{2}})$
    \citep{bakos:2010:hat11}.
}
\tablenotetext{d}{
    Values for a quadratic law, adopted from the tabulations by
    \cite{claret:2004} according to the spectroscopic (SPC) parameters
    listed in \reftabl{stellar}.
}
\tablenotetext{e}{
    The 95\% confidence upper-limit on the eccentricity based on the
    free-eccentricity model,
    with the primary constraint being the requirement that the
    observations match to a stellar evolution model. If we do not
    require a match to the stellar evolution models, then the 95\%
    confidence upper-limit on the eccentricity is $e < 0.125$,
    constrained by the RVs.
}
\tablenotetext{f}{
    Error term, either astrophysical or instrumental in origin, added
    in quadrature to the formal RV errors for the listed
    instrument. This term is varied in the fit assuming a prior inversely 
    proportional to the jitter.
}
\tablenotetext{g}{
    Correlation coefficient between the planetary mass \mpl\ and
    radius \rpl\ determined from the parameter posterior distribution
    via $C(\mpl,\rpl) = <(\mpl - <\mpl>)(\rpl -
    <\rpl>)>/(\sigma_{\mpl}\sigma_{\rpl})>$ where $< \cdot >$ is the
    expectation value operator, and $\sigma_x$ is the standard
    deviation of parameter $x$.
}
\tablenotetext{h}{
    Planet equilibrium temperature averaged over the orbit, calculated
    assuming a Bond albedo of zero, and that flux is reradiated from
    the full planet surface.
}
\tablenotetext{i}{
    The Safronov number is given by $\Theta = \frac{1}{2}(V_{\rm
    esc}/V_{\rm orb})^2 = (a/\rpl)(\mpl / \mstar )$
    \citep[see][]{hansen:2007}.
}
\tablenotetext{j}{
    Incoming flux per unit surface area, averaged over the orbit.
}
\ifthenelse{\boolean{emulateapj}}{
  \end{deluxetable*}
}{
  \end{deluxetable}
}

\subsection{Stellar Rotation}
\label{sec:rotation}

\begin{figure*}[]
\plotone{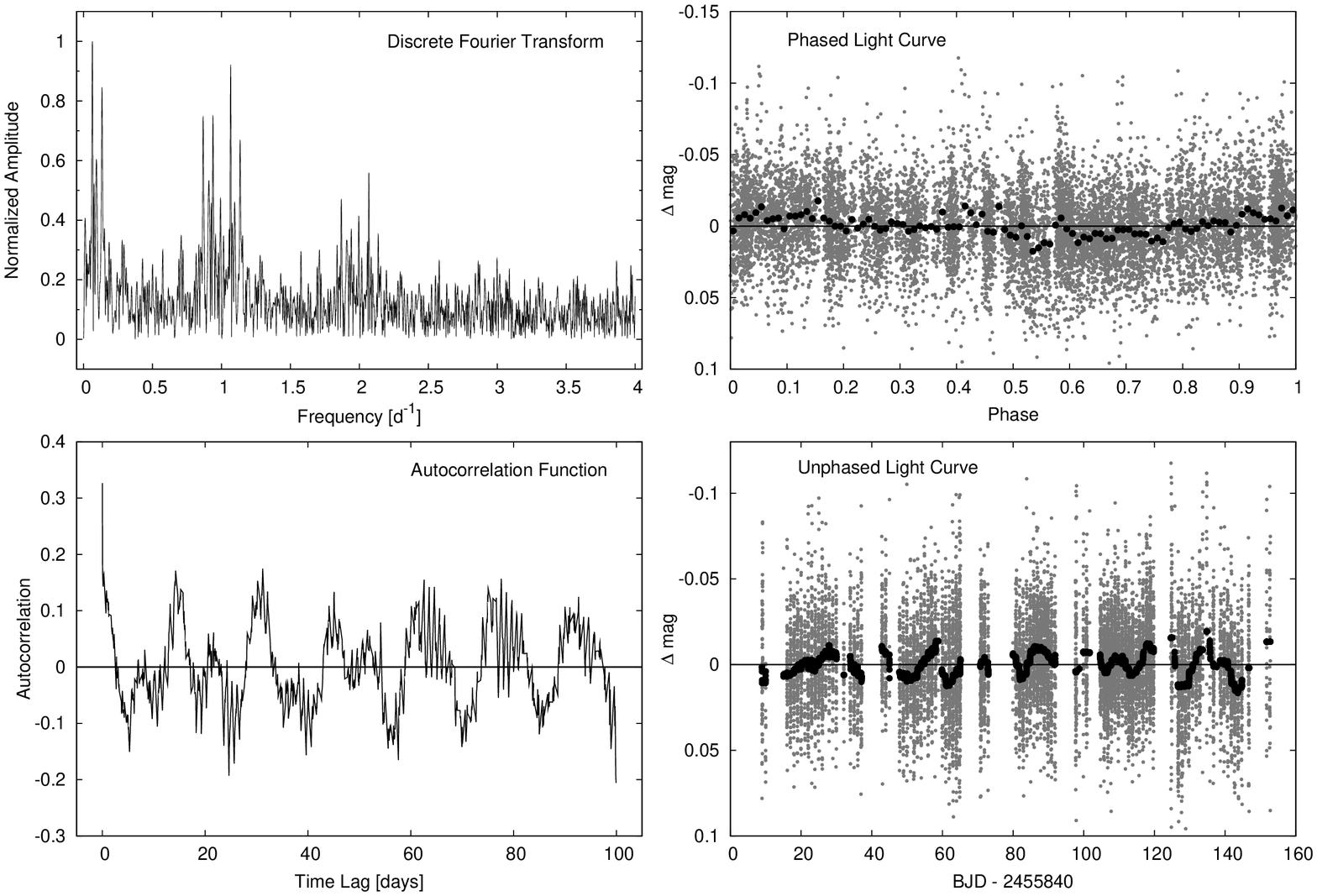}
\caption[]{
    Detection of quasi-periodic photometric variability in the HATNet
    \lc{} of \hatcur, which we attribute to spots on the surface of the
    rotating star.  Upper left: the Discrete Fourier Transform of the
    EPD-filtered HATNet \lc{} of \hatcur.  The amplitude has been
    normalized to that of the highest peak, seen at a frequency of $f =
    0.06397$\,d$^{-1}$ ($P = 15.6$\,d), and with an $S/N$ of 12.5.  We
    identify this as the photometric rotation frequency of the star. 
    Lower left: the Discrete Autocorrelation Function of the same light
    curve, calculated following \cite{edelson:1988}, and using a step
    of 0.05\,d.  The first significant peak seen at a lag of $15.6$\,d,
    corresponds to the rotation period.  The peak repeats with
    comparable amplitude through six cycles, indicating a long
    coherence time for the signal.  Upper right: the differential
    HATNet \lc{} phase-folded at the photometric rotation period of the
    star.  The light grey points show the individual measurements, the
    dark filled circles show the light curve binned in phase using a
    bin size of 0.01.  The solid line shows $\Delta m = 0$ for
    reference.  Lower right: un-phased light curve plotted as a
    function of time.  In this case the dark filled circles show the
    median light curve (calculated using a moving window of $1.0$\,d). 
    Variations with a peak to peak amplitude of $\sim 0.01$\,mag are
    apparent.
\label{fig:rotation}}
\end{figure*}

A search for continuous periodic variability in the residual HATNet
light curve (i.e., residuals after subtracting our model transit light
curve from the observations) using the Discrete Fourier Transform (DFT)
reveals a signal at a frequency of $f = 0.064247$\,d$^{-1}$.  This
signal is suppressed in the TFA light curve, but detected with a S/N of
12.5 and an amplitude of $5.6$\,mmag when only EPD is applied to the
light curve.  We tentatively identify this periodicity as the
photometric rotation frequency of the star.  The effective stellar
rotation period in that case is $P = 15.6$\,d, which is close to four
times the orbital period of the transiting planet.  The EPD residual
light curve phase-folded at this period, together with the DFT
spectrum, and the Discrete Autocorrelation Function
\citep[DACF;][]{edelson:1988} of the light curve are shown in
\reffigl{rotation}.  As seen in the DACF, the signal maintains
coherence through at least six cycles.  We note that the suppression by
TFA of relatively low-frequency ($f < 0.1$\,d$^{-1}$) stellar
variability due to rotation was previously seen in our analysis of
HAT-P-11 \citep{bakos:2010:hat11}, where a $P \sim 30$\,d signal was
found in the EPD HATNet light curve, but not found in the TFA light
curve.  Subsequent {\em Kepler} observations confirmed the variability
seen in the HATNet EPD data.

As a consistency check we may also estimate an upper limit on the
equatorial rotation period (assuming $\sin i = 1$) using the
spectroscopically determined $\vsini$ together with $\rstar$ as
determined in Section~\ref{sec:planetparam}.  We find $P <
13.1^{+3.6}_{-2.4}$\,d which is consistent with the photometric
rotation period to within $1\sigma$.

\section{Discussion}
\label{sec:discussion}

\begin{figure*}[]
\plotone{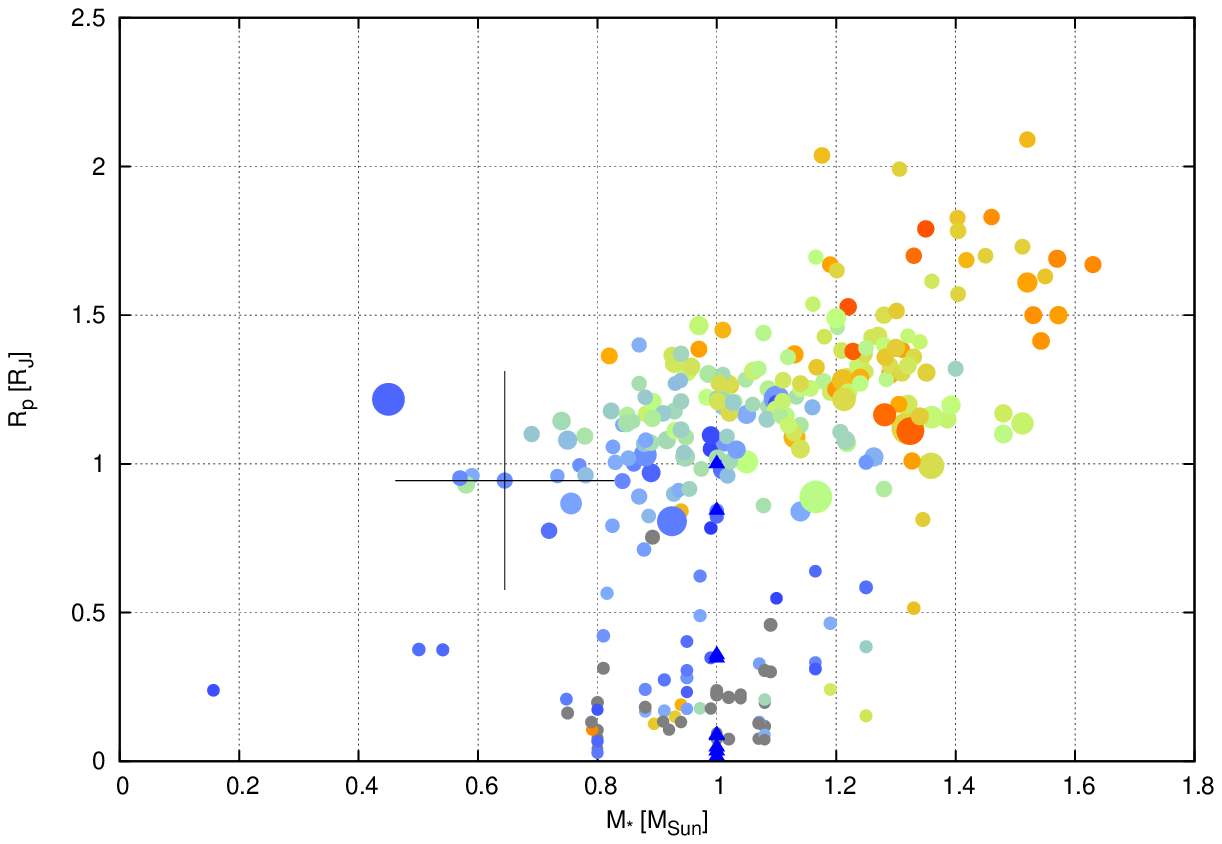}
\caption[]{
    Planetary radius versus stellar mass for transiting extrasolar
	planets (and brown dwarfs) with well measured masses (from radial velocities or
	transit timing variations).  Size of the points scales with the
	cube-root of the mass.  In the color version of this figure, color
	indicates equilibrium temperature of the planet (Bond albedo fixed
	at 0, full redistribution of flux assumed) on a rough scale of red
	being $T_{\rm eq} > 2000$\,K, green $\approx 1500$\,K and blue
	$\approx 1000$\,K.  Small blue triangles show the major Solar
	System planets.  HAT-P-54b is marked with a large cross.
\label{fig:mstar_rpl}}
\end{figure*}

In this paper we have presented the discovery and characterization of
\hatcurb{}, a compact hot Jupiter orbiting a late K dwarf star.  With a
mass of between that of Saturn and Jupiter (\hatcurPPmlong\,\mjup), and
radius of \hatcurPPrlong\,\rjup, \hatcurb{} is compact; smaller in
radius than $\sim 92$\% of the known transiting planets with measured
masses greater than that of Saturn.  \hatcurb{} also orbits one of the
lowest mass stars (\mstar = \hatcurISOmshort\,\msun) known to have a close-in
gas-giant planet, and it is the lowest mass planet host discovered by
the HATNet survey.  Only 3 stars with smaller mass are known to host a
planet with $0.1 < \mpl < 13$\,\mjup, and $P < $30\,d.  These are
WASP-43 \citep{hellier:2011}, WASP-80 \citep{triaud:2013}, and
Kepler-45 \citep{johnson:2012}.  A planet or brown dwarf with $\mpl =
18.4$\,\mjup\ has also been discovered on a 1.3\,d period orbit around
the M dwarf HD~41004~B \citep{zucker:2004}.  Thus, the \hatcur{} system
has several atypical properties which make it an interesting object for
further study.  These findings are well demonstrated by
\reffig{mstar_rpl}, displaying the planetary radius versus the
stellar mass.  \hatcur{} is shown by a cross at the left (small stellar
mass) and low (small planetary radius) side of the hot Jupiter
population.  Among a few other things worth noting in
\reffig{mstar_rpl} is the correlation of planet radius with
stellar mass, which is a manifestation of the previously found planet
radius vs.~stellar flux dependence \citep[see ][ and references
therein]{spiegel:2013}.  Also visible is the paucity of ``sub-Jovian''
planets with radii $0.4\,\rjup \lesssim \rpl \lesssim 0.7\,\rjup$, as
previously noted by \citet{beauge:2013} using the period--planetary
radius plane.  Finally, the emerging population of highly inflated
Jupiters \citep{hartman:2012:hat39hat41} also appears detached from the
rest of the planets on the top right side of the figure.

Of particular interest is the fact that \hatcur{} lies within field 0
of the K2 mission.  Observations of this field began on 2014 Mar 8, and
are expected to finish on 2014 May 30.  A proposal to observe this star
in long cadence mode, together with other candidate transiting planet
hosts identified by HATNet, has been accepted through the {\em Kepler}
Guest Observing program.  As far as we are aware, \hatcur{} is the only
currently known transiting planet within this field.  Several other
transiting planets are known near the K2 field, however none of them
are among the list of targets selected for observations with K2, and
all are outside the field defined by the selected 
targets\footnote{The
	list of targets proposed by the public and selected for observations
	can be found at
	http://keplerscience.arc.nasa.gov/K2/Fields.shtml\#0}. The
RV-detected planet HD~50554b \citep{fischer:2002}, which is not known
to be transiting, is also in field 0 of the K2 mission. Very high
precision {\em Kepler} observations of \hatcur{} will provide a wealth
of information that will be used to precisely measure the system
parameters, and possibly detect subtle features, such as spot
crossings by the transiting planet, and reflected light, among others.


\acknowledgements 

\paragraph{Acknowledgements}
HATNet operations have been funded by NASA grants NNG04GN74G and
NNX13AJ15G.  Follow-up of HATNet targets has been partially supported
through NSF grant AST-1108686.  G.\'A.B., Z.C.~and K.P.~acknowledge
partial support from NASA grant NNX09AB29G.  K.P.~acknowledges support
from NASA grant NNX13AQ62G.  G.T.~acknowledges partial support from NASA
grant NNX14AB83G.  We acknowledge partial support also from the Kepler
Mission under NASA Cooperative Agreement NCC2-1390 (D.W.L., PI). 
G.K.~thanks the Hungarian Scientific Research Foundation (OTKA) for
support through grant K-81373.  Based in part on observations obtained
at the W.~M.~Keck Observatory, which is operated by the University of
California and the California Institute of Technology.  Keck time has
been granted by NASA (N133Hr).  Data presented in this paper are based
on observations obtained at the HAT station at the Submillimeter Array
of SAO, and the HAT station at the Fred Lawrence Whipple Observatory of
SAO.  Data are also based on observations with the Fred Lawrence
Whipple Observatory 1.5\,m and 1.2\,m telescopes of SAO.  The authors
wish to recognize and acknowledge the very significant cultural role
and reverence that the summit of Mauna Kea has always had within the
indigenous Hawaiian community.  We are most fortunate to have the
opportunity to conduct observations from this mountain.

\clearpage
\bibliographystyle{apj}
\bibliography{htrbib.bib}

\end{document}